\documentclass[%
 aip,
 jmp,%
 amsmath,amssymb,
 reprint,%
 unsortedaddress
]{revtex4-1}

\usepackage{graphicx}
\usepackage{dcolumn}
\usepackage{bm}
\usepackage{multirow}
\usepackage{lmodern} 

\begin{document}

\preprint{AIP/123-QED}

\title{Steering Orbital Optimization out of Local Minima and Saddle Points Toward Lower Energy}

\author{Alain C. Vaucher}
\author{Markus Reiher}
\email[Corresponding author: ]{markus.reiher@phys.chem.ethz.ch}
\affiliation{ 
ETH Z\"urich, Laboratorium f\"ur Physikalische Chemie, Vladimir-Prelog-Weg 2, CH-8093 Z\"urich, Switzerland
}

\date{\today}

\begin{abstract}
The general procedure underlying Hartree--Fock and Kohn--Sham density functional theory calculations consists in optimizing orbitals for a self-consistent solution of the Roothaan--Hall equations in an iterative process.
It is often ignored that multiple self-consistent solutions can exist, several of which may correspond to minima of the energy functional.
In addition to the difficulty sometimes encountered to converge the calculation to a self-consistent solution, one must ensure that the correct self-consistent solution was found, typically the one with the lowest electronic energy.
Convergence to an unwanted solution is in general not trivial to detect and will deliver incorrect energy and molecular properties, and accordingly a misleading description of chemical reactivity.
Wrong conclusions based on incorrect self-consistent field convergence are particularly cumbersome in automated calculations met in 
high-throughput virtual screening, structure optimizations, \textit{ab initio} molecular dynamics, and in real-time explorations of chemical reactivity,
where the vast amount of data can hardly be manually inspected.
Here, we introduce a fast and automated approach to detect and cure incorrect orbital convergence, which is
especially suited for electronic structure calculations on
sequences of molecular structures. Our approach consists of a randomized perturbation of the converged electron density (matrix) intended to push SCF convergence to
orbitals that correspond to another stationary point (of potentially lower electronic energy) in the variational parameter space of an electronic
wave function approximation.
\end{abstract}

\maketitle

\setlength{\parindent}{0cm}
\setlength{\parskip}{0.6em plus0.2em minus0.1em}

\section{Introduction}

The rapid increase of computational power and the appearance of new technologies in the last decades have been a driving force in the development of algorithms in quantum chemistry.
This has facilitated the development of novel approaches such as real-time quantum chemistry,\cite{marti2009,haag2011,haag2013,haag2014a,haag2014b} the interactive optimization of molecular structures,\cite{bosson2012} interactive \textit{ab initio} molecular dynamics,\cite{luehr2015a} the interactive combination of atomistic simulations and art,\cite{davies2016a,mitchell2016a} new approaches to the visualization of quantum mechanical data,\cite{salvadori2016a} and the introduction of computer games in the related field of quantum physics.\cite{sorensen2016a,maniscalco2016a}

Underlying real-time quantum chemistry\cite{haag2013,haag2014b} is the ultra-fast quantum chemical calculation of molecular quantities in such a way
that they can be analyzed instantaneously.
We developed a computer program that allows one to interact with molecules during a quantum chemical exploration of sequences of molecular structures.\cite{haag2014b,vaucher2016a,muehlbach2016a,vaucher2016b}
Our program displays a three-dimensional visualization of a molecular system whose energy, gradients, and other properties are calculated continuously from first principles.
An operator can manipulate the molecular system by moving atoms or groups of atoms and will be able to immediately experience the effect of this manipulation on the molecular system.
In particular, our program can be coupled to a force-feedback haptic device that allows the operator to move atoms in three dimensions and instantaneously feel the force on these atoms brought about by the structure distortion.
The immediate (haptic and visual) feedback provides intuitive insights into the physics and chemistry of the molecular system under study.

We detected in some real-time reactivity explorations, molecular behavior that contradicted chemical knowledge 
(see below for examples) and that could be traced back to wrongly converged quantum chemical calculations.
The problem can be attributed to the existence of multiple self-consistent field (SCF) solutions in Hartree--Fock and Kohn--Sham density-functional theory (DFT) calculations.
Both theories require the solution of the Roothaan--Hall equations, which can be fulfilled by different self-consistent electronic densities \textit{for the same state and structure}.
While some solutions are unstable, others correspond to different energy minima in the space of the molecular orbital coefficients, of which in
general only the global minimum corresponds to the desired SCF solution.
Nevertheless, the integrity of such calculations is rarely verified, since usually the desired solution is obtained.
However, convergence to different SCF solutions is more common than expected.
It may occur and remain undetected in ordinary quantum chemical calculations. 

It is of utmost importance to ensure correct convergence of SCF calculations, especially in \textit{ab initio} molecular dynamics\cite{marx2009},
in interactive\cite{haag2014a,haag2014b} or automated\cite{Maeda2013,Zimmerman2013a,Rappoport2014,Bergeler2015,Zimmerman2015,Habershon2015,Habershon2016} explorations of chemical reactivity, and in high-throughput screening calculations\cite{bib:Hachmann1,bib:Lilienfeld5,bib:Rabie1,bib:Husch1,bib:Rinderspacher,bib:Lilienfeld1},
where results can hardly be manually validated.
Section~\ref{sec:scf_solutions} describes the origin and consequences of multiple SCF solutions in Hartree--Fock and Kohn--Sham DFT theory.
Then, Section~\ref{sec:scf_dynamic_simulations} puts this in the context of simulations involving consecutive electronic structure calculations for different geometries of a molecular system and illustrates the problem with some examples from real-time reactivity explorations.
Section~\ref{sec:solution} introduces a strategy to solve the problem and presents our implementation in a real-time quantum chemistry framework.
Then, in Sections~\ref{sec:examples} and~\ref{sec:comparison}, we apply the approach proposed in different real-time reactivity explorations and in standard quantum chemical calculations.

\section{Multiple SCF solutions in Hartree--Fock and Kohn--Sham DFT theories}\label{sec:scf_solutions}

In Hartree--Fock and Kohn--Sham DFT theories, one seeks to obtain the molecular orbitals that minimize the total energy (variational principle).
To find this energy minimum, one usually sets the condition of stationarity upon changes in the molecular orbitals, which delivers a set of non-linear equations.
These equations are a necessary, but not a sufficient condition to obtain the global energy minimum:\cite{adams1962a} one may obtain some solution which is a stationary point not necessarily the global or even a local energy minimum of the energy functional.

The central step underlying Hartree--Fock and Kohn--Sham DFT calculations is the iterative solution of the non-linear Roothaan--Hall equations, 
\begin{align}
  \boldsymbol{F} \boldsymbol{C} = \boldsymbol{S} \boldsymbol{C} \boldsymbol{\varepsilon},
\label{eq:roothaan-hall}
\end{align}
with the Fock matrix $\boldsymbol{F} = \boldsymbol{F}(\boldsymbol{C})$, the matrix of the molecular orbital coefficients $\boldsymbol{C}$, the overlap matrix $\boldsymbol{S}$, and the diagonal matrix $\boldsymbol{\varepsilon}$ containing the single-particle energies.
The molecular orbitals characterized by $\boldsymbol{C}$ are updated in every SCF iteration step, and the occupied ones enter the calculation of the electronic density or density matrix according to the Aufbau principle.
The electronic density (matrix) then defines the Fock matrix for the next iteration.
Unless degenerate orbitals are not equally occupied, the electronic density obtained at any iteration step is well defined.
However, the result of the whole SCF procedure depends on the initial Fock matrix (more precisely, on the molecular orbitals with which it is built) as well as on the convergence algorithm.

We emphasize that the multiple self-consistent solutions discussed here all fulfill the Aufbau principle and correspond to the same 
molecular structure and electronic state characterized by some global quantum numbers.
They may, however, correspond to different electronic energies.
For exact DFT, convergence to the correct ground-state density can be guaranteed for the exact ensemble functional.\cite{wagner2013a}

Note that we not only address unstable solutions (solutions that do not correspond to minima of the energy functional):
we also consider that different SCF solutions may correspond to minima of the energy functional, of which the lowest one is of interest.

While different SCF solutions are characterized by different energies, a Hartree--Fock or Kohn--Sham DFT energy should be unique.
To avoid ambiguities in the calculation of Hartree-Fock or Kohn--Sham DFT energies, 
only one SCF solution can be the correct one (except for degenerate solutions).
One is generally interested in the solution that delivers the lowest electronic energy, which is sometimes called the global\cite{deandrade2005a,thom2008a} or absolute\cite{paldus1970a} minimum, or Hartree--Fock ground state.\cite{adams1962a}
Here, we will call it global minimum.
Usually, the established quantum chemical algorithms find this solution directly.

In quantum chemical calculations, the choice of quantum numbers and symmetries 
determines which solutions are obtained, but the fundamental problem of multiple SCF solutions will remain.
For example, multiple local minima have been reported for closed-shell Hartree--Fock calculations\cite{thom2008a} and Perdew--Zunger self-interaction corrected DFT calculations relying on complex orbitals.\cite{lehtola2016a}
In the following, the discussion and calculations refer to unrestricted calculations, but our results have 
analogous implications for restricted calculations or for calculations with complex orbitals.
It can be argued whether and when applying the spin-unrestricted formalism and choosing one of its solutions allows for an adequate description of the 
electronic wave function.\cite{pulay1988a,pulay1990a,jimenezhoyos2011a,small2015b,toth2016a}
Nevertheless, especially in Kohn--Sham DFT studies, one usually relies on such wave functions\cite{frenking2000a,ziegler2005a,neese2009a,jacob2012a} 
although for a single-determinant approximation a symmetry must be broken
(total spin\cite{lykos1963a,noodleman1979a,noodleman1981a,noodleman1982a,jacob2012a} or particle number\cite{scuseria2011a}) 
in order to deal with certain static correlation cases.
Furthermore, unrestricted orbitals even allow for compact expansions of full configuration interaction calculations.\cite{thomas2014a}

\section{Series of similar molecular structures}\label{sec:scf_dynamic_simulations}

\subsection{Multiple SCF solutions for consecutive calculations of similar molecular structures}

In real-time quantum chemistry, \textit{ab initio} molecular dynamics, and geometry optimizations, electronic structure calculations are executed for many consecutive molecular structures.
In general, the electronic densities for two consecutive structures are very similar, since the coordinates of the atomic nuclei change only moderately between two single-point calculations.
As a consequence, a converged density matrix is an adequate guess for the next step, which also reduces computational time considerably.
This can further be improved by extrapolation techniques.\cite{atsumi2008,atsumi2010,muehlbach2016a}

Along a reaction path, however, there may be discontinuities or bifurcations with respect to the correct SCF solution.
At such points, the previous density matrix is not an adequate guess and may lead to convergence to some local-minimum SCF solution.
In such cases, wrong quantum chemical properties are obtained (especially, incorrect nuclear forces), which in turn steer the whole process into 
a wrong direction.

\subsection{Examples of incorrect convergence for contiguous sequences of structures}\label{sec:preliminary_examples}

In the following, we illustrate for simple examples how, in a sequence of molecular structures, quantum chemical results, which are initially correct,
lead to incorrect chemical behavior.
In the first three examples, the electronic density is caught in an initially correct restricted solution (in an unrestricted framework).
The fourth example shows that undesired convergence can also occur in open-shell cases with odd numbers of electrons.
The sequences of structures of this section were generated by real-time manipulation of molecules in our real-time quantum chemistry framework.\cite{haag2014b,vaucher2016a,muehlbach2016a,vaucher2016b}
Note that sequences generated in automated calculations (such as structure optimizations)
or in molecular dynamics simulations could also bring about similarly incorrect behavior.

In a real-time setting, odd chemical behavior is easily detectable by experiencing unexpected atomic forces when operating a 
haptic device or by observing unexpected bond orders (calculated, for instance, according to Mayer\cite{mayer1983} and indicated in the 
visualization by an automatically adjusted width of the sticks indicating chemical bonds).

All energies and forces reported for the following examples were obtained from unrestricted calculations with the PM6 method.\cite{stewart2007}
The one-dimensional exploration coordinate in all figures below that report PM6 calculations is given without units. 
However, we note that the numbers assigned to the exploration coordinate in
these figures actually denote seconds and correspond to the time required for the real-time exploration.
The explicit Cartesian coordinates along such a collective coordinate are provided in the Supporting Information.

\subsubsection{Hydrogen abstraction in CH$_4$}

\begin{figure}[htb]
\centering
\includegraphics[width=0.35\textwidth]{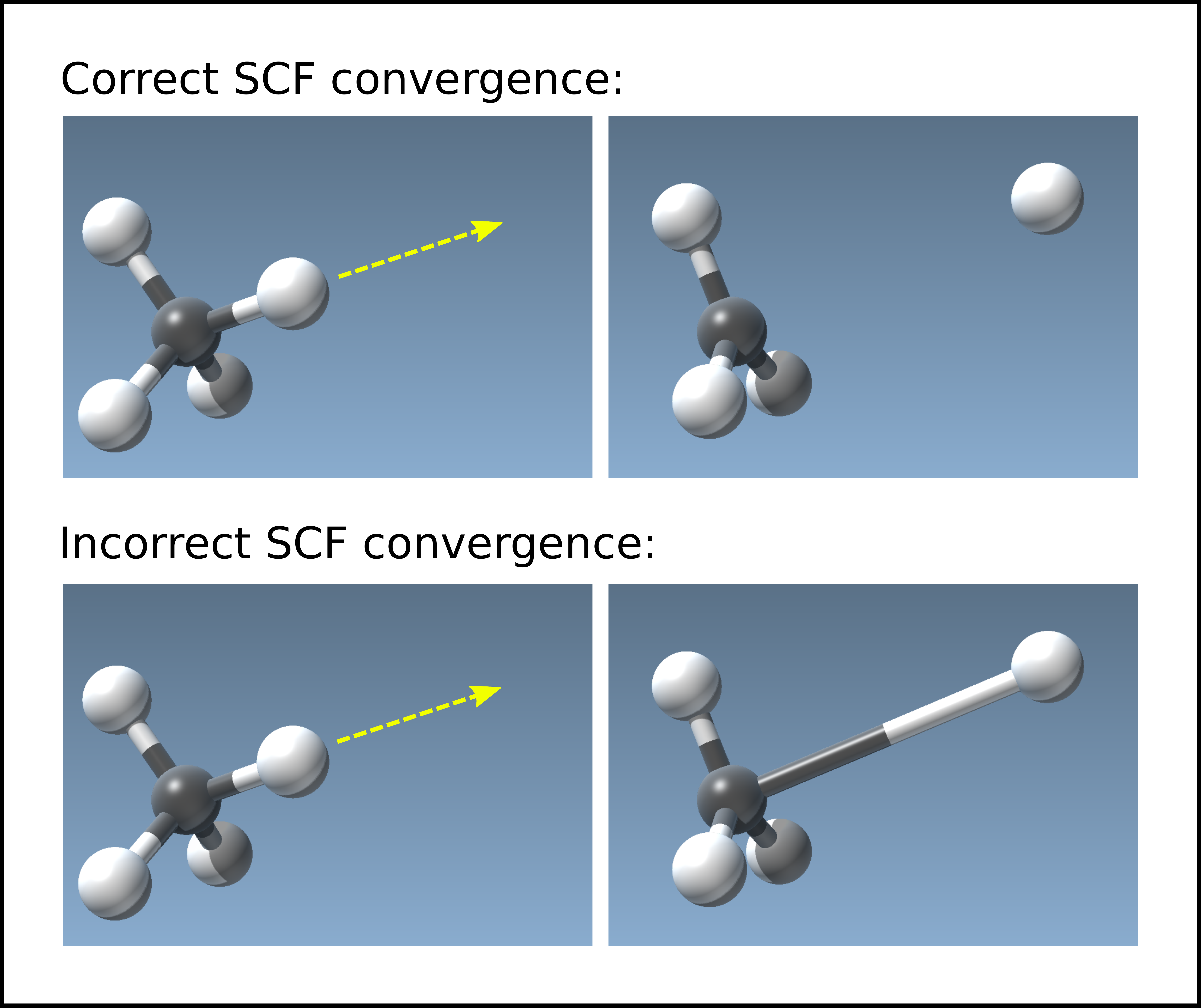}
\caption{
  Abstraction of a hydrogen atom in methane.
  Top: Correct, expected behavior.
  Bottom: Incorrect behavior realized when no special care is taken regarding SCF convergence and
flagged by a non-vanishing Mayer bond order between the abstracted hydrogen atom and the carbon atom (indicated by a stick representing the bond). 
  Hydrogen and carbon atoms are depicted in white and black, respectively.
}
\label{fig:ch4}
\end{figure}

A well-known example for undesired SCF convergence is the hydrogen abstraction in methane (Fig.~\ref{fig:ch4}).
In the equilibrium geometry, the correct SCF solution is of (spin-)restricted character.
Upon dissociation of one of the hydrogen atoms, the system remains in a restricted solution, even when the underlying calculation is based on the unrestricted formalism.
A spin-symmetry breaking in an unrestricted calculation can eventually produce the lowest-energy $\mathrm{H}^\cdot + {}^\cdot\mathrm{CH}_3$ dissociation 
channel in a single-determinant framework, but it can be necessary to enforce this as otherwise the restricted (electron-pairing) solution may prevail.

In a real-time exploration of this hydrogen abstraction reaction with a haptic device, the operator will feel a strong attraction towards the carbon atom 
even after pulling the hydrogen atom to a considerable distance.
Furthermore, the visualization points to wrong orbital convergence as
a bond is drawn because of a Mayer bond order close to one between the abstracted hydrogen atom and the carbon atom (Fig.~\ref{fig:ch4}), 
whereas none would be expected.

\subsubsection{Hydrogen atom manipulation in ethane}

\begin{figure}[htb]
\centering
\includegraphics[width=0.45\textwidth]{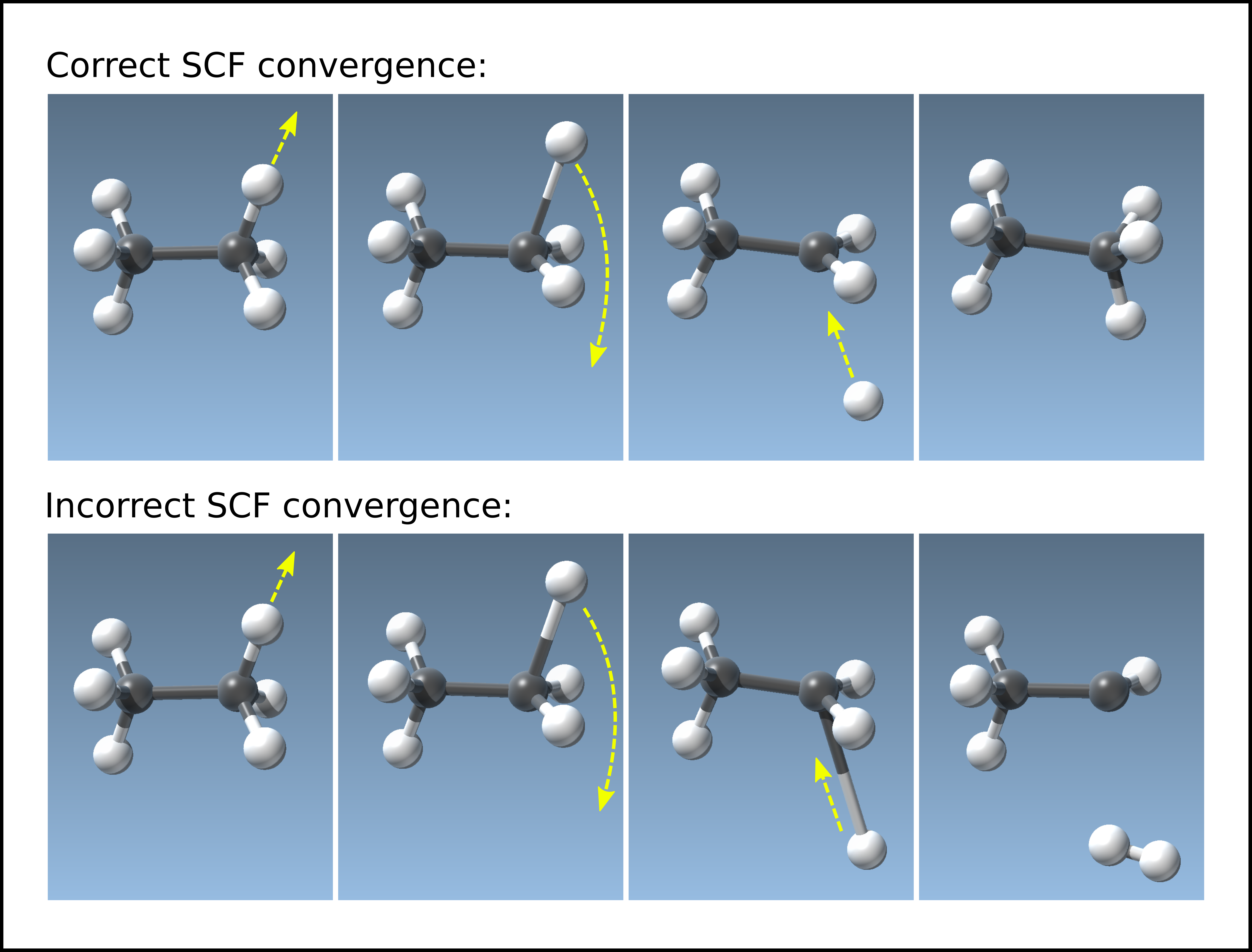}
\caption{
  Abstraction and re-binding of a hydrogen atom in ethane.
  Top: Correct behavior.
  Bottom: Incorrect behavior, realized when no special care is taken regarding SCF convergence.
  Hydrogen and carbon atoms are depicted in white and black, respectively.
}
\label{fig:c2h6}
\end{figure}

In a real-time manipulation of one of the hydrogen atoms of ethane, transferring it from one side of the molecule to the other may result in 
artificial atomic forces (repulsion from the carbon atom) and eventually in dihydrogen dissociation, instead of regenerating an ethane molecule in a different conformation.
This is illustrated in Fig.~\ref{fig:c2h6}.

\subsubsection{Rotation around the double bond of ethene}

\begin{figure}[htb]
\centering
\includegraphics[width=0.4\textwidth]{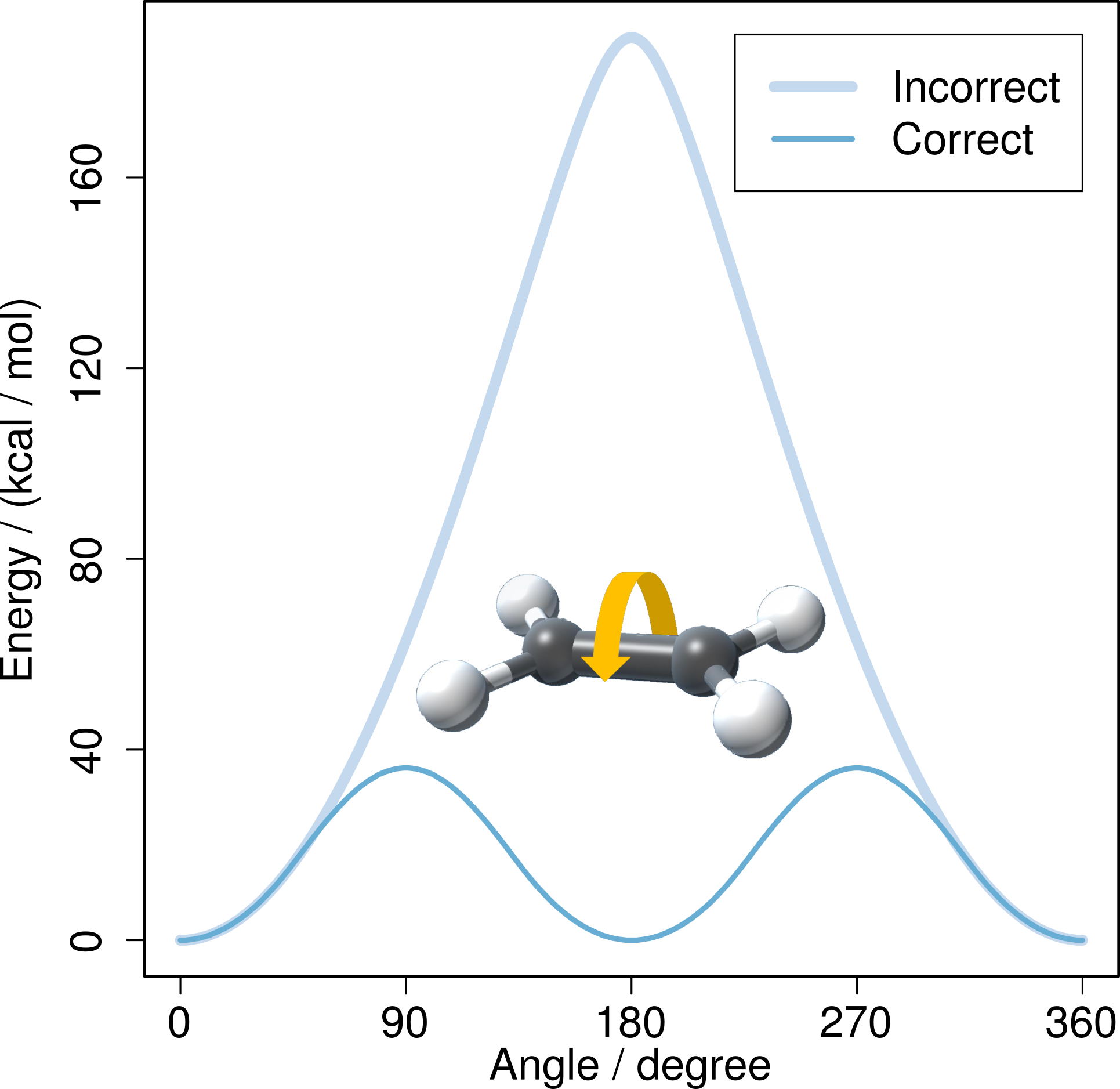}
\caption{
  PM6 energy profiles for the rotation around the double bond of ethene.
  The incorrect profile is obtained when no special care is taken regarding SCF convergence and each converged electronic density is a guess for the next molecular structure.
  Hydrogen and carbon atoms are depicted in white and black, respectively.
}
\label{fig:c2h4}
\end{figure}

Fig.~\ref{fig:c2h4} shows energy profiles for a rotation around the double bond of ethene.
One expects to obtain an energy minimum for angles of 0, 180, and 360 degrees, which correspond to identical structures, as well as an energy maximum for angles of 90 and 270 degrees.
However, consecutive calculation of the structures in the sequence may deliver an incorrect energy profile with an energy maximum at 180 degrees.

\subsubsection{Radical polymerization of ethene}

\begin{figure}[h!]
\centering
\includegraphics[width=0.4\textwidth]{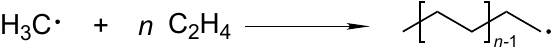}
\caption{Model reaction for ethene polymerization.}
\label{fig:radpol}
\end{figure}

\begin{figure}[htb]
\centering
\includegraphics[width=0.4\textwidth]{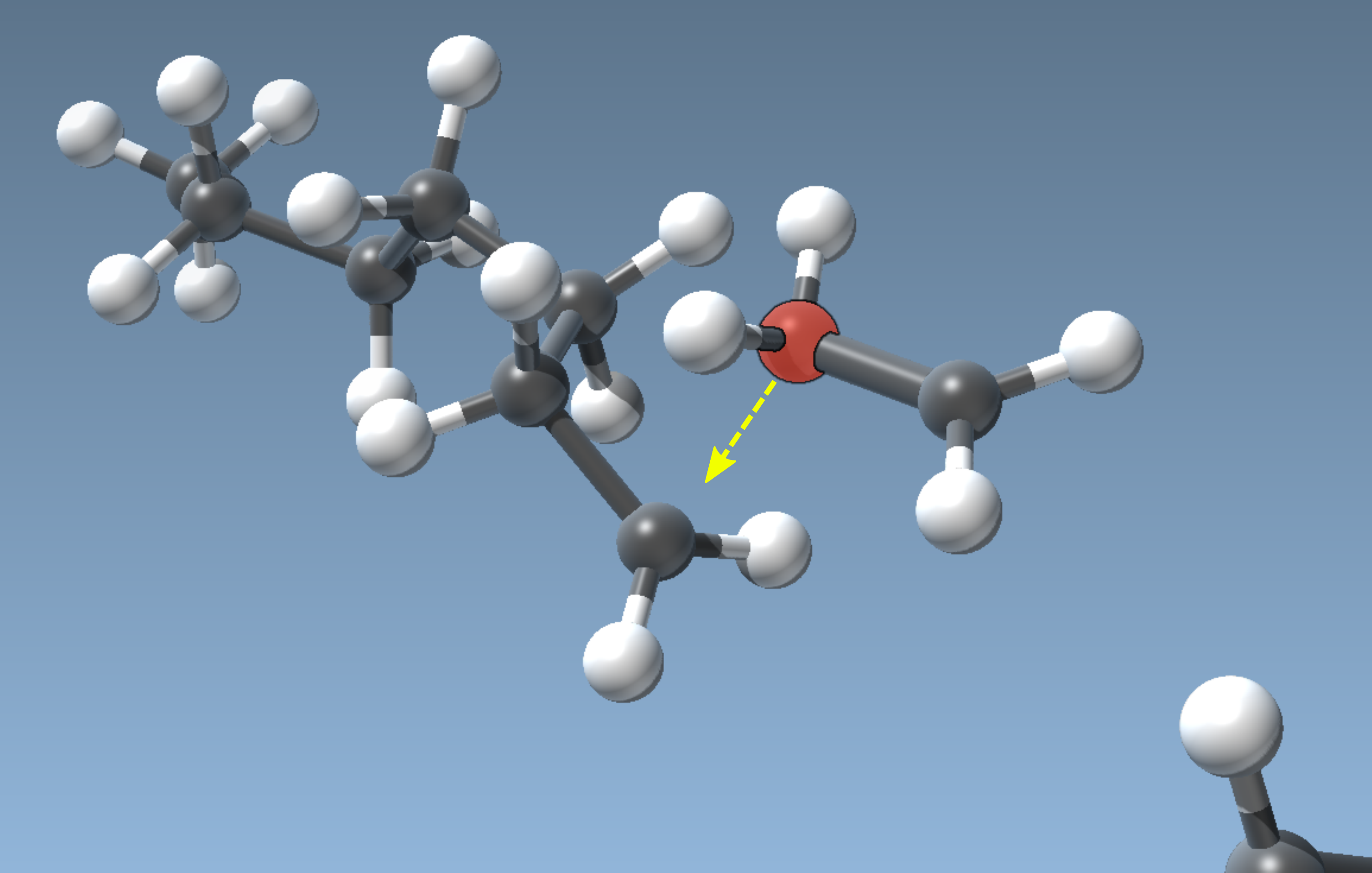}
\caption{Cumbersome step during an interactive exploration of the radical polymerization of ethene.
  The carbon atom depicted in red is moved towards the radical end of the chain with a haptic device, but the radical end moves out of the way and no reaction occurs.
  Hydrogen and carbon atoms are depicted in white and black, respectively.
}
\label{fig:polymerization_screenshot}
\end{figure}

The next example considers ethene polymerization starting from a methyl radical (Fig.\ \ref{fig:radpol}).
In a real-time exploration of this reaction, it is sometimes not possible to make the ethene molecule react with the polymer 
radical by moving it towards the growing chain with the haptic device (Fig.~\ref{fig:polymerization_screenshot}):
one experiences a repulsive force, and the radical end of the polymer moves away.

\section{Enforcing SCF convergence to lowest-energy solutions}\label{sec:solution}

\subsection{General considerations}

Many strategies exist to enable or improve SCF convergence into a self-consistent solution (see, for example, Ref.~\citenum{schlegel1991} and references cited therein).
However, as already observed more than 30~years ago, ``computational efforts are usually concentrated on convergence alone, and a solution that has converged in both the energy and density matrix is usually accepted as the `true' solution, within the given approximations''.\cite{mezey1980a}
It is seldom verified that the true solution (global SCF minimum) is obtained and not an unstable solution (such as a saddle point) or another local minimum.
Compared to the issue of convergence alone, little work has been done about locating the global SCF minimum.

Unstable SCF solutions can be detected by stability analyses.\cite{thouless1960a,adams1962a,cizek1967a,paldus1970b,cizek1970a,seeger1977a,bauernschmitt1996a}
A stability analysis consists in calculating the lowest eigenvalue of the electronic Hessian, which will be negative for an unstable solution.
Stability analysis is a very useful tool to verify that the obtained SCF solutions are minima of the energy functional and not just saddle points.
Although stability analyses can be applied to any Hartree--Fock or Kohn--Sham wave function, it is usually only applied to avoid getting stuck in a closed-shell solution.\cite{pulay1990a,yamada2015a,toth2016a} 
Accordingly, stability analyses often search for singlet and triplet instabilities. 
In comparison, unstable broken-symmetry solutions are rarely discussed.
Stability-analysis calculations are commonly not applied because of the relatively high computational cost necessary for the construction of the electronic Hessian.
Furthermore, stability analyses cannot distinguish between local and global minima.

To be certain that a SCF solution is the correct one, little can be done but to ensure that there exists no other self-consistent solutions with 
lower energy.
For this, one needs to explore thoroughly the space of possible electronic densities (density matrices), 
as in SCF metadynamics\cite{thom2008a} or in simulated annealing approaches.\cite{deandrade2005a,deandrade2006a}

In real-time quantum chemistry, convergence to non-global minima must be detected automatically and as rapidly as possible.
While only exhaustive search can ensure that the correct SCF solution has been found, this is impractical except for very small systems: 
the high dimensionality of the parameter space in which the solution must be found implies unaffordable computational cost for ultra-fast 
or big-data applications.

In the following, we present a solution for real-time quantum chemistry that is satisfactory in terms of speed and efficacy.
Afterwards, we discuss how to modify it for single-point calculations and \textit{ab initio} molecular dynamics.

\subsection{Strategy for real-time quantum chemistry}

Smoothness in real-time reactivity explorations is achieved by frequent delivery of nuclear forces.
In order to preserve this smoothness, the solution presented here involves no change in the calculations 
delivering the energies and forces underlying the exploration.
Instead, our approach consists in launching, in the background, frequent additional single-point calculations that attempt to find a lower energy for the last electronic structure calculation executed in the reactivity exploration.
Each of these calculations starts from a different electronic density (matrix) guess, and will converge either to the same SCF solution or to another one of higher or lower energy.

This guess for the additional calculation is obtained by perturbing the converged molecular orbitals of the reference single-point calculation.
The aim is to sufficiently modify the guess electronic structure in order to steer the SCF calculation into different self-consistent solutions.
It is important to randomize this perturbation.
Randomizing it allows us to explore consecutively different regions of the molecular orbital space and therefore increases the chance of reaching the basin leading to the global minimum if this solution is different from the one already found.
Furthermore, without randomization the perturbation would always deliver similar guesses, which is not optimal. 

We implement the perturbation by randomly selecting multiple pairs of occupied--unoccupied molecular orbitals and randomly mixing them.
We found that the selection of ten pairs works very well
for all cases studied here, but this may, of course, be changed on input. As our perturbation calculations are carried out for 
contiguous sequences of molecular structures, we found that it is not necessary to perturb every structure and that 
one such perturbation calculation per structure is sufficient, as a different
(randomly) perturbed guess will be tried for a similar structure shortly after the last perturbation calculation. Clearly, for individual 
structures one may inspect more than one perturbed guess at a time.

Our perturbation approach can be compared to the one applied in the SCF metadynamics implementation of \textsc{Q-chem},\cite{shao2015a}
and we note that other perturbation schemes are also possible. By contrast,
electron smearing (see, e.g., Refs.\ \cite{merm65,meth89,degi95,vers01,basiuk2011a}) allows for a fractional occupation of orbitals and 
is usually applied for orbitals that are difficult to converge because of 
close lying frontier orbitals. In general, however,
it does not help to detect incorrect SCF convergence. Our approach, however, introduces more
drastic changes to the guess orbitals in order to drive convergence to other minima in parameter space if possible.

If a perturbation calculation finds the same SCF solution or one with a higher energy, it will be ignored.
However, if a SCF solution with a lower energy is obtained, the corresponding molecular orbitals will be injected 
into the reactivity exploration as a guess density matrix for the next single-point calculation.
The injection of another electronic density introduces a discontinuity in observables assigned to the sequence of structures.
For ultra-fast calculations, this may be acceptable, but it can be avoided
by reseting the calculation to previous structures of the reactivity exploration.

\begin{figure*}[htb]
\centering
\includegraphics[width=0.95\textwidth]{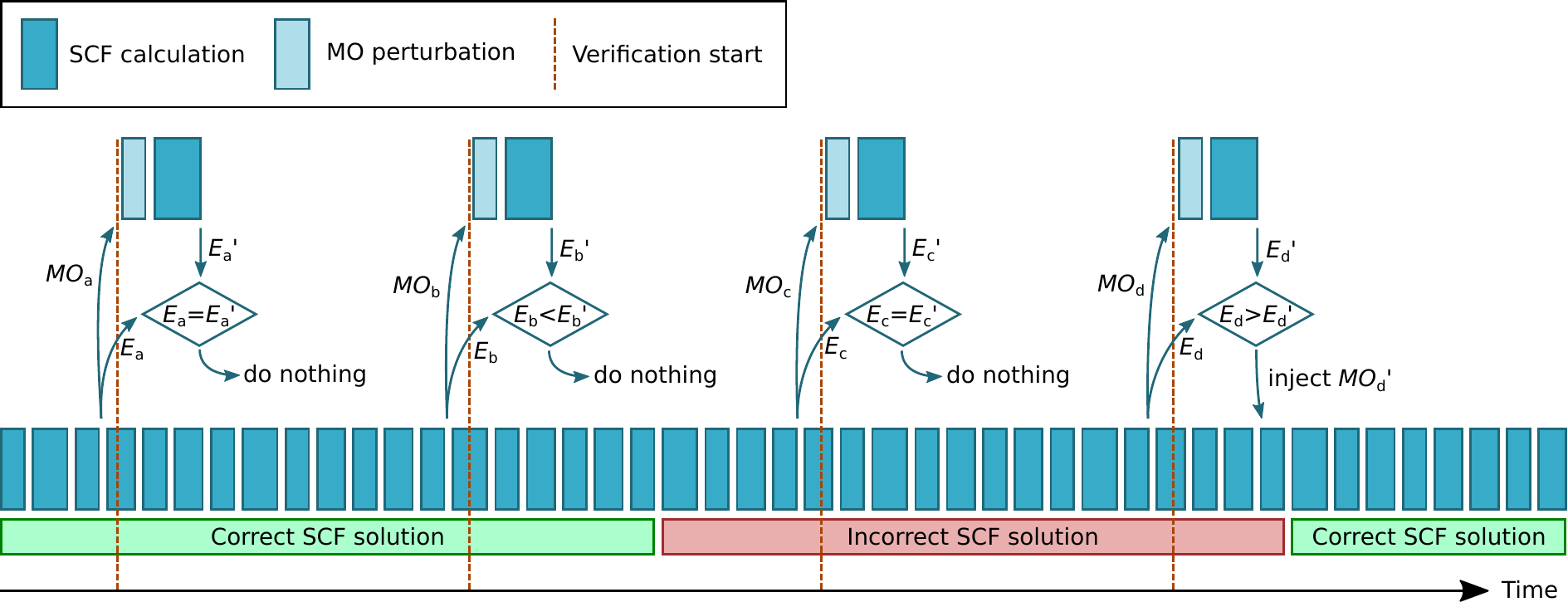}
\caption{
  Schematic representation of our correction scheme for wrong convergence behavior in real-time quantum chemistry.
  The quantum chemical properties needed for the real-time reactivity exploration are continuously delivered by SCF calculations in the main calculation thread, represented by the blue boxes in the lower half of the figure.
  In real-time reactivity explorations, the execution time for such SCF calculations is typically on the order of a few milliseconds.
  Thereby, from some point onwards they converge to an incorrect SCF solution.
  In regular intervals, convergence verifications are executed in the background by re-executing a SCF calculation, starting from perturbed molecular orbitals (MO).
  Here, four verifications are displayed. 
  The first two (\textit{a} and \textit{b}) cannot detect better convergence.
  The third verification (\textit{c}) does not detect that the convergence of the main calculation thread is incorrect because it finds the same incorrect SCF solution.
  However, the fourth verification (\textit{d}) finds a SCF solution with a lower energy and, accordingly, the correct molecular orbitals are injected as a guess for the next SCF calculation.
}
\label{fig:strategy}
\end{figure*}

This strategy is illustrated schematically in Fig.~\ref{fig:strategy}.
It cannot guarantee that an incorrect SCF convergence will be detected immediately.
However, in a real-time setting the additional calculations are very fast, so that they can be performed frequently without hampering the reactivity exploration.
In case of incorrect convergence, this strategy will end up finding the correct solution with a high probability.
In general, the many verifications, each of which starts from different electronic densities, ensure that this happens very rapidly (see Section~\ref{sec:examples} for examples).

\subsection{Implementation}\label{sec:implementation}

We implemented  approach was implemented in our real-time framework, which is interfaced to the \textsc{Samson} molecule viewer.\cite{samson050}

For the perturbation of the molecular orbitals, our current scheme selects randomly, out of the 15 highest occupied and the 15 lowest unoccupied molecular orbitals, 10 random pairs, which are mixed with a random angle between 0 and 90 degrees in such a way that the randomly selected pair of orbitals $\phi_\mathrm{o}$ (occupied) and $\phi_\mathrm{v}$ (unoccupied) are superimposed with the random angle $\alpha$ according to 
\begin{align}
  \phi_\mathrm{o,new} = \cos{\alpha} \cdot \phi_\mathrm{o,old} + \sin{\alpha} \cdot \phi_\mathrm{v,old} \\
  \phi_\mathrm{v,new} = \cos{\alpha} \cdot \phi_\mathrm{v,old} - \sin{\alpha} \cdot \phi_\mathrm{o,old}
\end{align}
(recall that this procedure is repeated in regular intervals for similar structures of a contiguous series such that
a good coverage of the relevant parameter space is achieved).
An angle $\alpha$ of 0 degrees corresponds to no mixing at all and an angle $\alpha$ of 90 degrees swaps the orbitals.
The additional calculations for the convergence verification are carried out five times per second.

When incorrect convergence is detected, this is recorded in a log file in addition to injecting the correct orbitals into the main exploration.
The convergence verifications are activated by default and can be switched off if necessary.

\subsection{Single-point calculations, structure optimizations, and \textit{ab initio} molecular dynamics}\label{sec:approach_variants}

The orbital perturbation introduced here to detect and eventually avoid undesired orbital convergence can also be beneficial
in single-point calculations, structure optimizations, and \textit{ab initio} molecular dynamics.
For single-point calculations, a computer program can, after converging a set of orbitals, launch one or several calculations after 
perturbation of the converged molecular orbitals as discussed above. It can then automatically inspect whether a solution with lower energy is found.
While this cannot ensure that the solution is the global minimum, it is an inexpensive, efficient, and reliable test as will be seen in the examples below.
Such a test is complementary to standard stability analysis and increases the confidence that a SCF solution is neither a saddle point nor a non-global minimum.
For \textit{ab initio} molecular dynamics, the orbital perturbations can be executed every few steps of a simulation trajectory.
The same is possible for structure optimizations.
Should an incorrect convergence be detected, the calculations can be rewound to the point where the bifurcation occurred.
Although this will increase the computation time, it will be worth the effort and probably unavoidable if simple manual
inspection is not possible (e.g., in cases of a vast number of calculations or for complex chemical processes for which one does
not have any expectations or prior understanding).

\section{Examples}\label{sec:examples}

The approach presented in Section~\ref{sec:solution} solves the incorrect behavior in all examples of Section~\ref{sec:preliminary_examples}.
Out of these examples we describe the polymerization reaction in detail below. In addition, two examples are presented for which 
it is not trivial to spot insufficient SCF convergence during a reactivity exploration.

For all examples, the energies were obtained with the PM6 method.\cite{stewart2007}
The interested reader can find videos for the real-time explorations of the six examples of Section~\ref{sec:preliminary_examples} and of this Section on the internet.\cite{convergenceVideos}
Below, the profiles obtained with and without convergence verification are compared.
The corresponding coordinate files can be found in the Supporting Information.
In all cases where we observed a bifurcation of energy profiles due to our orbital perturbation approach
this was detected almost instantaneously so that the exploration required no rewinding to previous structures.

\subsection{Ethene polymerization}

When exploring the radical polymerization of ethene starting from the methyl radical (discussed
in Section~\ref{sec:preliminary_examples}) with convergence verification as described above, the reaction can be reproduced as expected.

\begin{figure}[htb]
\centering
\includegraphics[width=0.45\textwidth]{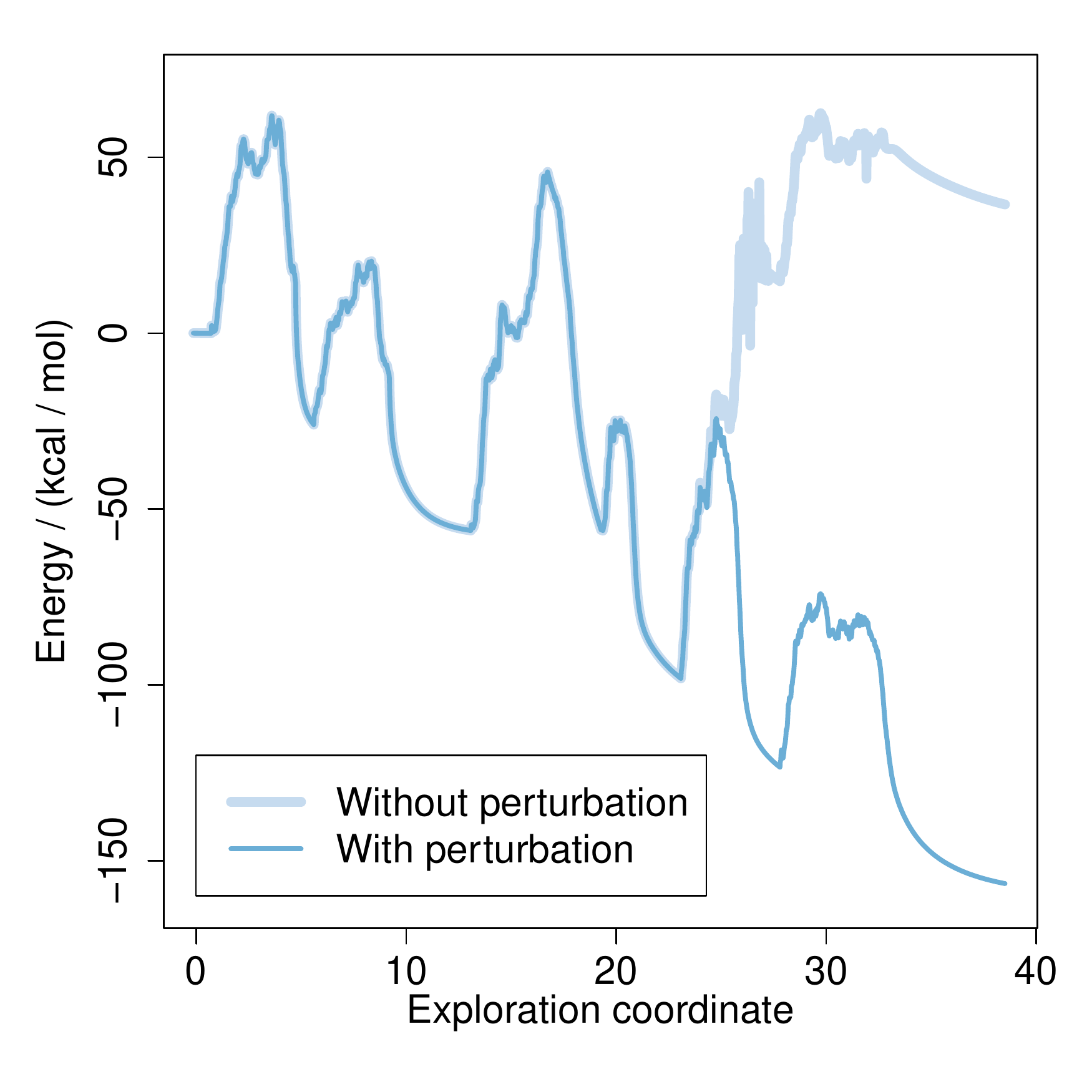}
\caption{
  Electronic PM6 energy for ethene polymerization in a real-time exploration.
  The energy barriers represent the consecutive additions of ethene molecules to the growing chain.
  Note that the recorded trajectory does not correspond to a minimal energy path.
}
\label{fig:polymerization_profile}
\end{figure}

Fig.~\ref{fig:polymerization_profile} shows the energy profile for the recorded trajectory with and without convergence verification.
When adding the fifth ethene molecule, the profile without convergence verification starts to depart from the one obtained by our
orbital perturbation calculations.

\subsection{Corey--Chaykovsky epoxidation}

\begin{figure}[h!]
\centering
\includegraphics[width=0.35\textwidth]{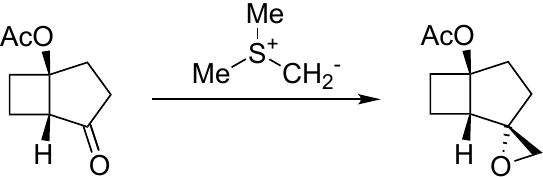}
\caption{
  Corey--Chaykovsky epoxidation.
  }
\label{fig:epoxidation_reaction}
\end{figure}

In a previous work on molecular propensity,\cite{vaucher2016b} we studied the Corey--Chaykovsky epoxidation shown in Fig.~\ref{fig:epoxidation_reaction} in our real-time quantum chemistry framework.

\begin{figure}[htb]
\centering
\includegraphics[width=0.45\textwidth]{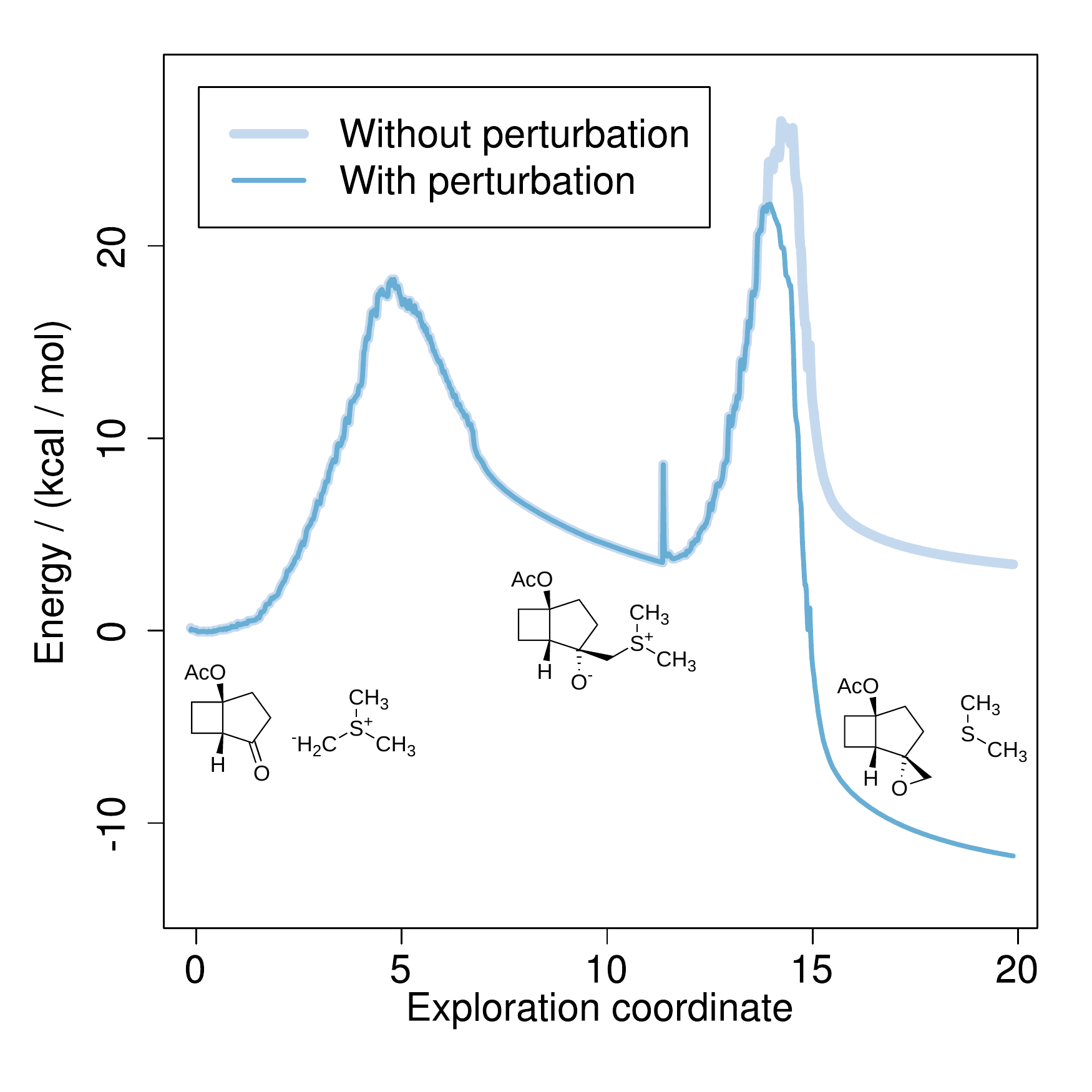}
\caption{
  Electronic PM6 energies for the Corey--Chaykovsky epoxidation.
  The peak around coordinate 11.5 appeared when the operator chose to change the position of the carbon atom to carry out
  the second part of the reaction.
}
\label{fig:corey_profile}
\end{figure}

For the reaction path taken there, the energy profile will run into a wrong SCF solution if no special care is taken, as shown in Fig.~\ref{fig:corey_profile}.
As the spin symmetry of the incorrect profile is already broken here, the origin for the incorrect convergence is not triplet instability.
We note that we were already aware of the convergence issue in our previous work\cite{vaucher2016b} and the energies reported there are correct.

\subsection{Oxidation of hydrogen by the iron oxide cation}

\begin{figure}[htb]
\centering
\includegraphics[width=0.45\textwidth]{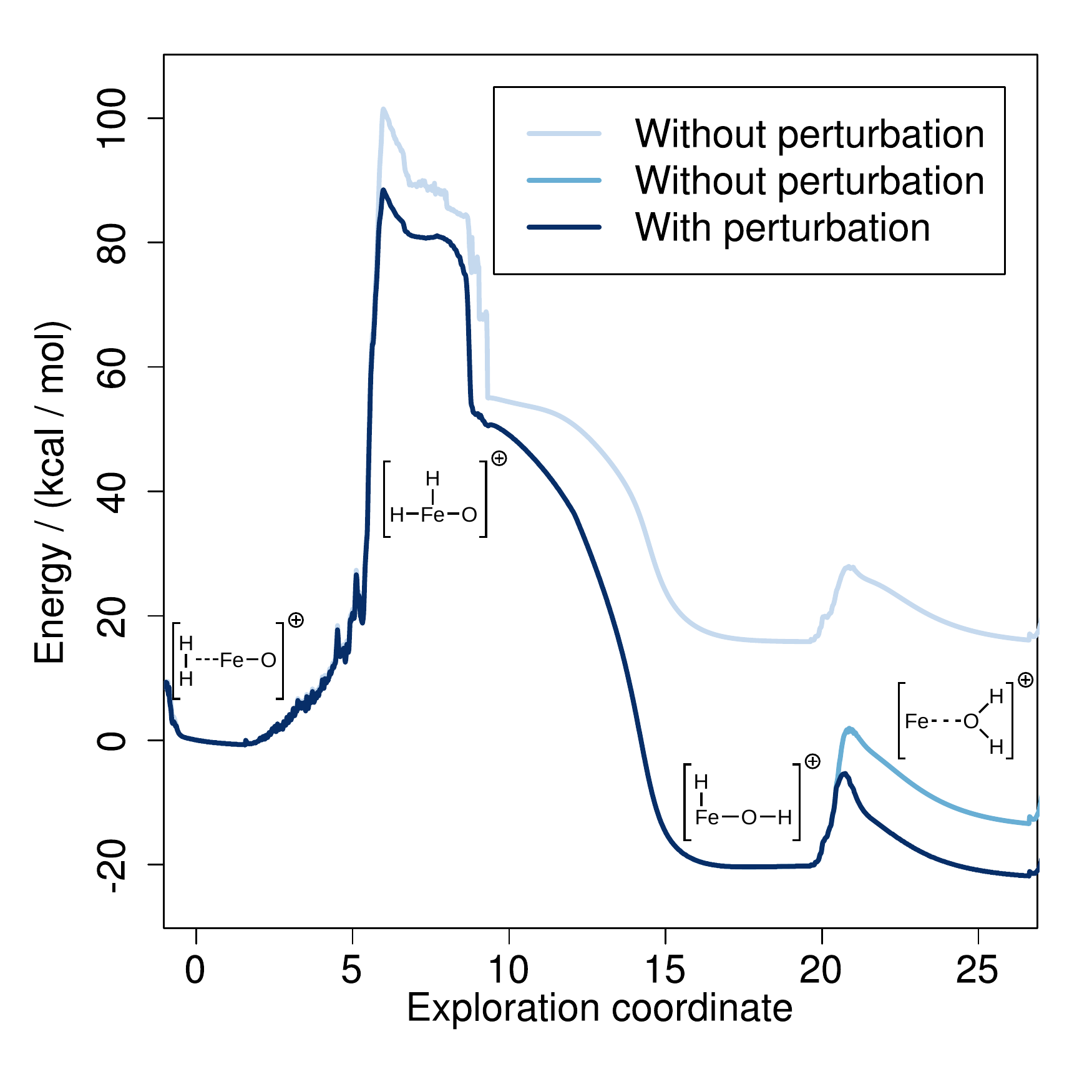}
\caption{
  Electronic PM6 energies for the oxidation of hydrogen by the iron oxide cation (lowest quartet spin state).
  The second incorrect profile was obtained when resetting the correct molecular orbitals at the intermediate.
}
\label{fig:iron_profile}
\end{figure}

The reaction path for 
\begin{equation*}
	\text{FeO}^+ + \text{H}_2 \longrightarrow \text{Fe}^+ + \text{H}_2\text{O}
\end{equation*}
is also taken from our work on molecular propensity.\cite{vaucher2016b}
For this reaction, both the lowest-lying sextet and quartet spin states are of interest.
When no special care is taken with regard to convergence, the sextet profile is correct (as one might have expected), 
but in the quartet state the calculations run into an incorrect SCF solution (Fig.~\ref{fig:iron_profile}).
We note again that the results reported in Ref.~\citenum{vaucher2016b} show the correct convergence behavior
and that our perturbation approach solves the problem automatically.

\section{Comparison with standard DFT calculations}\label{sec:comparison}

To show that the severe convergence issues discussed in this paper are a general problem for computational chemistry not limited to semi-empirical methods, two of the examples considered above were studied with standard DFT methods (PBE0\cite{adamo1999a}, def2-TZVP\cite{weigend2005}), for which we applied the program \textsc{Turbomole}\cite{ahlrichs1989} (version~6.5).
We chose this program for its very stable and reliable convergence acceleration implementation (as, for instance, witnessed in numerous
computational studies on open-shell transition-metal clusters reported by our group in the past fifteen years).

For both examples, we calculated three energy profiles. 
First, for each structure, we calculated the energies delivered when starting the calculation from an Extended-H\"{u}ckel Theory (EHT) guess.
Second, we calculated an energy profile where the starting orbitals for each structure are taken from the converged orbitals of the previous structure.
Finally, we added a perturbation of the molecular orbitals for every fifth structure and injected the corresponding molecular orbitals 
if the calculation had delivered a lower energy.
For this, we wrote a utility program that transforms the molecular orbitals delivered by \textsc{Turbomole} according to the algorithm presented in Section~\ref{sec:implementation}.
In all callculations we appplied the standard \textsc{Turbomole} settings, in particular a total electronic energy convergence criterion of
10$^{-6}$, unless otherwise noted.

\subsection{Rotation around the double bond of ethene}

\begin{figure}[htb]
\centering
\includegraphics[width=0.45\textwidth]{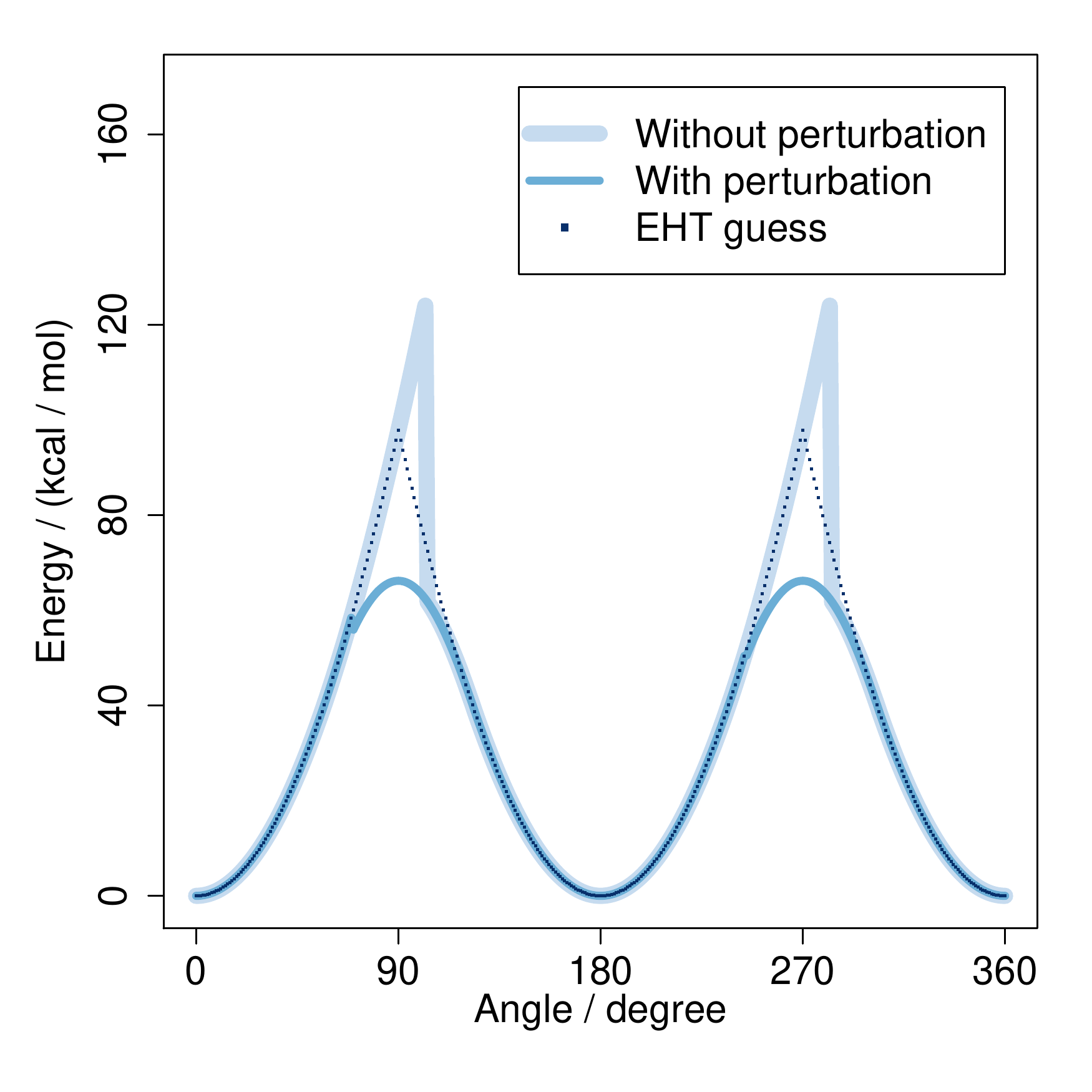}
\caption{
  Electronic PBE0/def2-TZVP energies for rotation around the double bond of ethene.
  The trajectory consists of 361 structures. 
  The perturbed energy profile is obtained by perturbing the molecular orbitals every fifth structure.
}
\label{fig:rotation_turbomole}
\end{figure}

The DFT energies obtained for the rotation around the double bond of ethene are illustrated in Fig.~\ref{fig:rotation_turbomole}.

Here, both the EHT guess and the normal propagation of the molecular orbitals deliver wrong energies for structures around angles of 90 and 270 degrees.
The energies obtained with the EHT guess reflect the restricted solution, and normal propagation of the orbitals runs into another restricted solution for angles slightly larger than 90 and 270 degrees, before finding the correct, unrestricted solution.
With the perturbation of the molecular orbitals, the simulation switches rapidly to the correct solution.

The three SCF solutions obtained at an angle of 102 degrees were analyzed in detail, and a table summarizing their differences is available in the Supporting Information.
There, isosurfaces for the highest occupied and lowest unoccupied molecular orbitals show that their electronic structure is considerably different.
For instance, the order of the orbitals in the two closed-shell solutions is different (although the Aufbau principle is fulfilled in both cases).
Stability analysis of the (unrestricted) SCF solutions obtained with the EHT guess or without perturbation reveal a triplet instability.
Recalculation of these unstable solutions in the restricted formalism and subsequent stability analysis of the closed-shell solutions show that they are local minima in the restricted formalism (i.e.~they have no singlet instability).

Interestingly, for the structure with an angle of 90 degrees, employing the triplet molecular orbitals as a guess for the singlet calculation 
still produces the same incorrect solution. In this case, setting a tighter convergence threshold for the total electronic energy 
10$^{-9}$ is necessary to find the correct solution after thousands of SCF iterations.

\subsection{Oxidation of hydrogen by the iron oxide cation}

\begin{figure}[htb]
\centering
\includegraphics[width=0.45\textwidth]{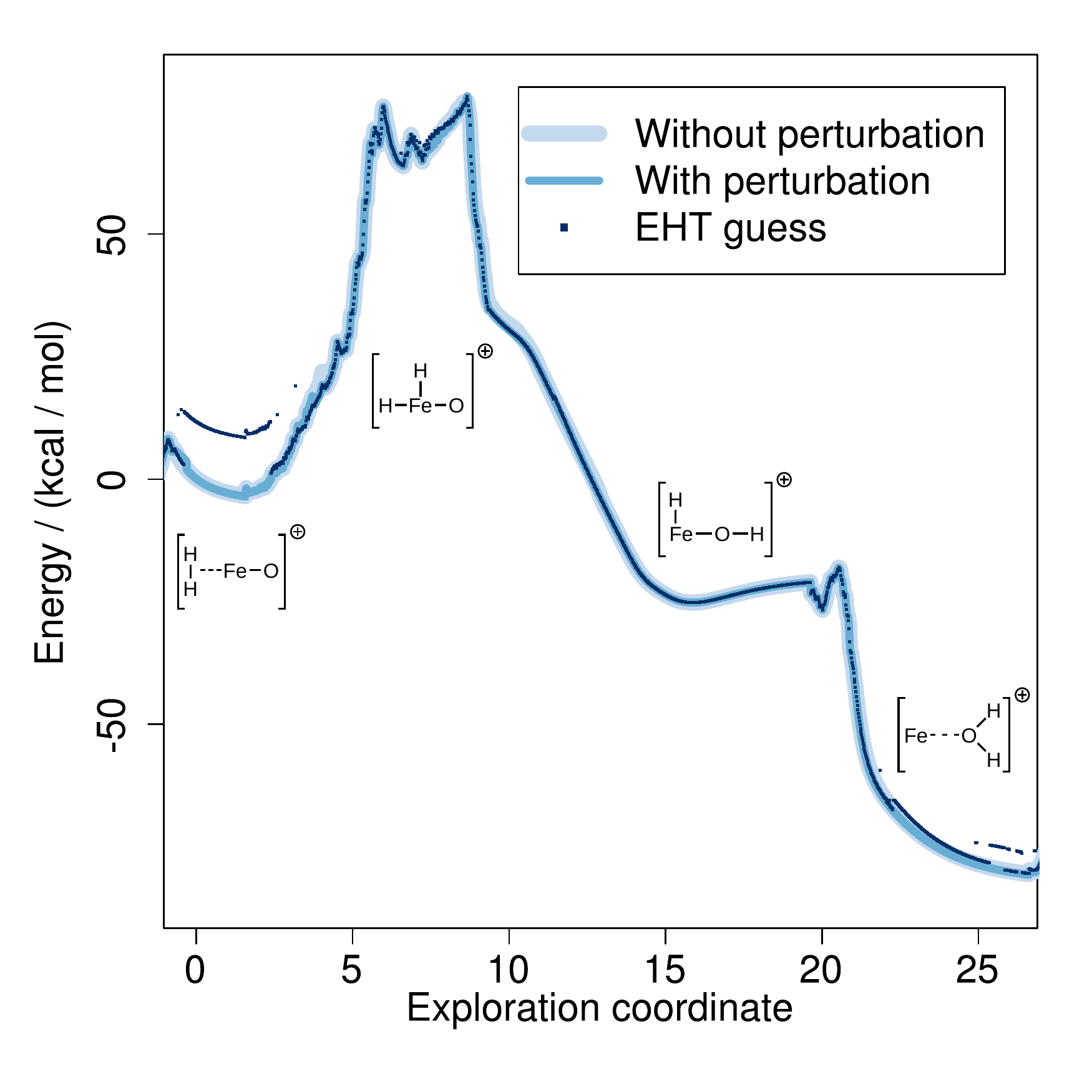}
\caption{
  Electronic PBE0/def2-TZVP energies for the oxidation of hydrogen by the iron oxide cation for the lowest quartet spin state.
  The trajectory displayed here consists of 1500 structures. 
  The perturbed energy profile is obtained by perturbing the molecular structures every fifth structure.
}
\label{fig:iron_turbomole}
\end{figure}

The three energy profiles obtained for the sextet state of the oxidation of hydrogen by the iron oxide cation are identical and will not be shown here.
However, for the quartet state, they all differ as can be seen in Fig.~\ref{fig:iron_turbomole}
(the Supporting Information contains enlarged figures for the sake of clarity).

The three energy profiles deviate little, but various observations can be made.
First, the energies obtained in this example confirm the existence of different SCF solutions.
Especially, we see that the solution resulting from the EHT guess is sometimes considerably different from
the lowest-energy solution obtained here.
Second, the energies of the propagated profile are largely correct, but sometimes correspond to a non-global minimum and slightly differ in the energy (for instance at coordinates around 3.8, 8.5, 9.5 and 12.5, see Supporting Information).
Third, the perturbed profile does not always deliver the lowest energy either. However, the perturbations allow for detecting this early, and, as a whole, the perturbed profile is better than both other profiles in this case.

A more detailed analysis reveals both stable and unstable incorrect SCF solutions.
For instance, at coordinate 25.5, three different energies are delivered by the three calculations.
While a stability analysis for the solution obtained from the EHT guess detects an instability (negative eigenvalue in the electronic Hessian), the other two solutions are stable.
After enforcing convergence of the unstable solution into a minimum, a fourth SCF solution is found, three of which are distinct, non-degenerate local SCF minima.
The Supporting Information contains a table summarizing the properties of these four SCF solutions.
Whereas some of them are characterized by qualitatively different molecular orbitals, others seem to be similar. In the case
of the latter, also the total electronic energies of the three local SCF minima are very similar. Still, they represent
different solutions as it is not possible to converge them to one solution by enforcing the convergence thresholds. Note that the
kinetic and potential energy contributions of these solutions with similar energies are notably different.


\section{Conclusions}

Many quantum chemical studies rely on Hartree--Fock and Kohn--Sham DFT theories.
An unpleasant feature of these theories is that they allow for multiple solutions.
In general, only one of them is of interest: the solution yielding the lowest electronic energy.
Usually, the established quantum chemical algorithms find the correct solution directly.
There exist, however, cases in which SCF calculations fail to find the correct electronic ground-state density and some other self-consistent solution is found.

When the quantum chemical calculations find incorrect SCF solutions, the calculated properties are unreliable.
In real-time quantum chemistry, incorrect SCF convergence can sometimes be spotted easily, for instance,
by experiencing highly artificial force feedback or because bonds are displayed where none is expected.

This problem cannot be solved trivially and so no simple and universal solution exists.
In particular, stability analysis will only be able to detect unstable self-consistent solutions of the energy functional (such as saddle points), but cannot provide information on whether a solution corresponds to a local or to the global minimum.

To avoid incorrect behavior and ensure the reliability of real-time reactivity explorations, the strategy adopted in this work 
is to continuously search, in the background, for a better SCF solution than the one currently delivering energy and forces underlying a simulation.
With drastic random perturbations of the molecular orbitals, these background convergence verifications detected 
incorrect SCF convergence very rapidly.
While our approach cannot guarantee with certainty that incorrect SCF convergence will be detected, we have discovered no case where it failed
and it has become a vital component of our real-time quantum chemistry framework. 

Since single-point calculations, structure optimizations, and \textit{ab initio} molecular dynamics can also suffer from 
incorrect SCF convergence, we discussed variants of our approach that can be beneficial in these settings.
From the examples shown in this work, it is evident that convergence to undesired solutions, be it saddle points or non-global minima of the energy functional, are more common than expected.
The potential benefit of applying the approach presented in this work outweights its extra computational cost, even in case of expensive SCF calculations.
In \textit{ab initio} molecular dynamics simulations and structure optimizations, such verifications are not necessary at every time or structure step so 
that the additional cost are managable.

Whereas this work focused on single-reference methods, similar convergence issue may arise 
in multi-configuration SCF (MC-SCF) calculations. MC-SCF methods also rely on non-linear equations and hence
MC-SCF solutions can, in principle, be caught in local minima.\cite{deandrade2005a}

\section*{Acknowledgments}
The authors gratefully acknowledge support by ETH Zurich (grant number: ETH-20 15-1).

\section*{References}

%

\end{document}